\documentclass[3p, twocolumn]{elsarticle}
\usepackage{ amsmath,amssymb, underscore}
\pdfoutput=1
\usepackage{subfig,float}
\usepackage{multicol}
\usepackage{graphicx}
\usepackage{color}
\usepackage{xspace}
\usepackage{hyperref}
\hypersetup{  pdfauthor={...},  pdftitle={...},  pdfsubject={...},  urlcolor=blue,}

\begin{document}

\renewcommand{\figurename}{Fig.}

\title{Reanalyzing the upper limit on the tensor-to-scalar perturbation ratio $r_T$ in a quartic potential inflationary model}

\author[du]{R. Kabir\corref{cor1}}
\ead{rakesh@physics.du.ac.in }

\author[du]{A. Mukherjee\corref{cor2}}
\ead{am@physics.du.ac.in }

\author[du]{D. Lohiya\corref{cor3}}
\ead{dlohiya@physics.du.ac.in }

\cortext[cor1]{Corresponding author}

\address[du]{Department of Physics and Astrophysics, University of Delhi, Delhi-110007, India}


\begin{abstract}

We study the polynomial chaotic inflation model with a single scalar field in a double well quartic potential which has recently been shown to be consistent with Planck data. In particular, we study the effects of lifting the degeneracy between the two vacua on the inflationary
observables, i.e., spectral index  $n_s$  and tensor-to-scalar perturbation ratio $r_T$. We find that removing the degeneracy allows the model to satisfy the upper limit constraints on $r_T$ from Planck data, provided the field starts near the local maximum. We also calculate the scalar power spectrum and non-Gaussianity parameter $f_{\text{NL} }$ for the primordial scalar perturbations in this model.

\end{abstract}

\maketitle

\section{ Introduction}

Inflation is regarded as the standard cosmological paradigm to describe the physics of the very early Universe. It leads to a causal mechanism to generate almost scale invariant fluctuations on cosmological scales, with small deviations that follow from the precise microphysics of inflation. This prediction is consistent with the recently announced measurements
of the cosmic microwave background (CMB) anisotropies by the Planck satellite. The latest data allow us to constrain the inflationary model besides giving a slightly red tilted spectral index \(n_s\) = 0.9603 $\pm $ 0.0073, ruling out exact scale invariance \(n_s\) = 1 at over 5$\sigma $ \cite{Ade2013}. The Planck data also provide observational bounds on primordial non-Gaussianity parameter, i.e., local
$ f_{\text{NL}} 
= 2.7 \pm  5.8 $ at 68\% confidence level for the reduced bispectrum \cite{Collaboration2013}.

Just after Planck results were announced, different scalar potentials were revisited to explore concordance with these results \cite{Ade2013}. The data reinforced the
ruling out of single-field inflationary models $V (\phi ) = \phi^{n}$ with $n \geq 2$, which were already disfavored or marginally disfavored
by WMAP. (This does not apply to alternative inflationary scenarios e.g., warm inflation \cite{ramospower2013}). However Croon et al. \cite{Croon2013} demonstrated that the double well degenerate potential $V (\phi ) =A \phi ^2 (v^2-2 v B\phi +\phi ^2)$ is consistent with Planck data, although with severe constraints on the initial condition for $\phi $ and on the allowed range of $B$. The polynomial quartic potential with double well is the most studied potential in a variety of settings. It is also well motivated by the physics beyond the Standard Model, viz. supergravity and superstring theories \cite{Nakayama2013}. The characteristic interesting features \cite{Croon2013, Jain2009} of the potential have been the primary motivation for a lot of  effort to explore its consistency with the Planck data  \cite{Ade2013}.

\section{ Basic Formalism}

\subsection{Slow Roll Parameters}

We review the formalism in \cite{Croon2013} to set up our notation. For an inflationary model described by the  potential $ V (\phi ) =A \phi^2 (v^2-2 v B \phi +\phi^2)$, the slow-roll parameters are defined as 

\begin{equation}
\epsilon  \equiv   \frac{1}{2}M_{\text{pl}}^2 \left(\frac{V_{\phi }}{V}\right)^2\text
= 2M_{\text{pl}}^2 \frac{\left(v^2-3 v B \phi +2 \phi
^2\right)^2}{\phi ^2 \left(v^2-2 v B \phi +\phi ^2\right)^2} \, , 
\end{equation}

\begin{equation}
\eta    \equiv M_{\text{pl}}^2 \left(\frac{V_{\phi \phi }}{V}\right) =\text2M_{\text{pl}}^2 \frac{ \left(v^2-6 v B \phi +6 \phi ^2\right)}{\phi
^2 \left(v^2-2 v B \phi +\phi ^2\right)}
\end{equation}
and
\begin{align}
\xi  & \equiv  M_{\text{pl}}^4 \left(\frac{V_{\phi }V_{\phi \phi \phi }}{V^2}\right)\\
 & =\text{  }24M_{\text{pl}}^4\text{  }\frac{ (-v B+2 \phi ) \left(v^2-3
v B \phi +2 \phi ^2\right)}{\phi ^3 \left(v^2-2 v B \phi +\phi ^2\right)^2} \notag
\end{align}
where
\begin{equation}
V_{\phi} \equiv \dfrac{dV}{d\phi},  V_{\phi\phi} \equiv \dfrac{d^2V}{d\phi^2} \text{ and }  V_{\phi\phi\phi} \equiv \dfrac{d^3V}{d\phi^3} \, .
\end{equation}

In terms of the slow-roll parameters, the scalar spectral index is expressed as
\begin{equation}
n_s= 1 - 6 \epsilon  + 2 \eta
\end{equation}

and the tensor-to-scalar ratio as
\begin{equation}
r_T= 16 \epsilon \, .
\end{equation}

Finally the number of e-folds is given by

\begin{equation}\label{Ne}
N = -\frac{1}{M_{\text{pl}}^2}\int_{\phi _i}^{\phi _e} (\frac{V}{V_{\phi }}) \, d\phi
\end{equation}
where \(\phi _{e, i }\) are the values of $\phi $ at the end and beginning of the inflationary epoch.

The scalar power spectrum $P_s(k)$ is described in terms of the spectral index  $n_{s }(k)$ by 
 \begin{equation}
 P_s(k) = A \exp [(n_{s }-1) \ln  (k /k_0) +\frac{1}{2} \alpha _s \ln ^2 (k/k_0)]
 \end{equation}
where
\begin{equation}
\alpha _s \equiv \frac{d n_s}{d \ln k}
\end{equation}
which in terms of slow-roll parameters is
\begin{equation}
\alpha_s= \frac{1}{8 \pi ^2}[-\frac{\xi  }{4} + 2\eta \epsilon  - 3\epsilon ^2] \, .
\end{equation}

\begin{figure}[H]
\begin{center}
\quad
\subfloat[]{\includegraphics[width=75mm]{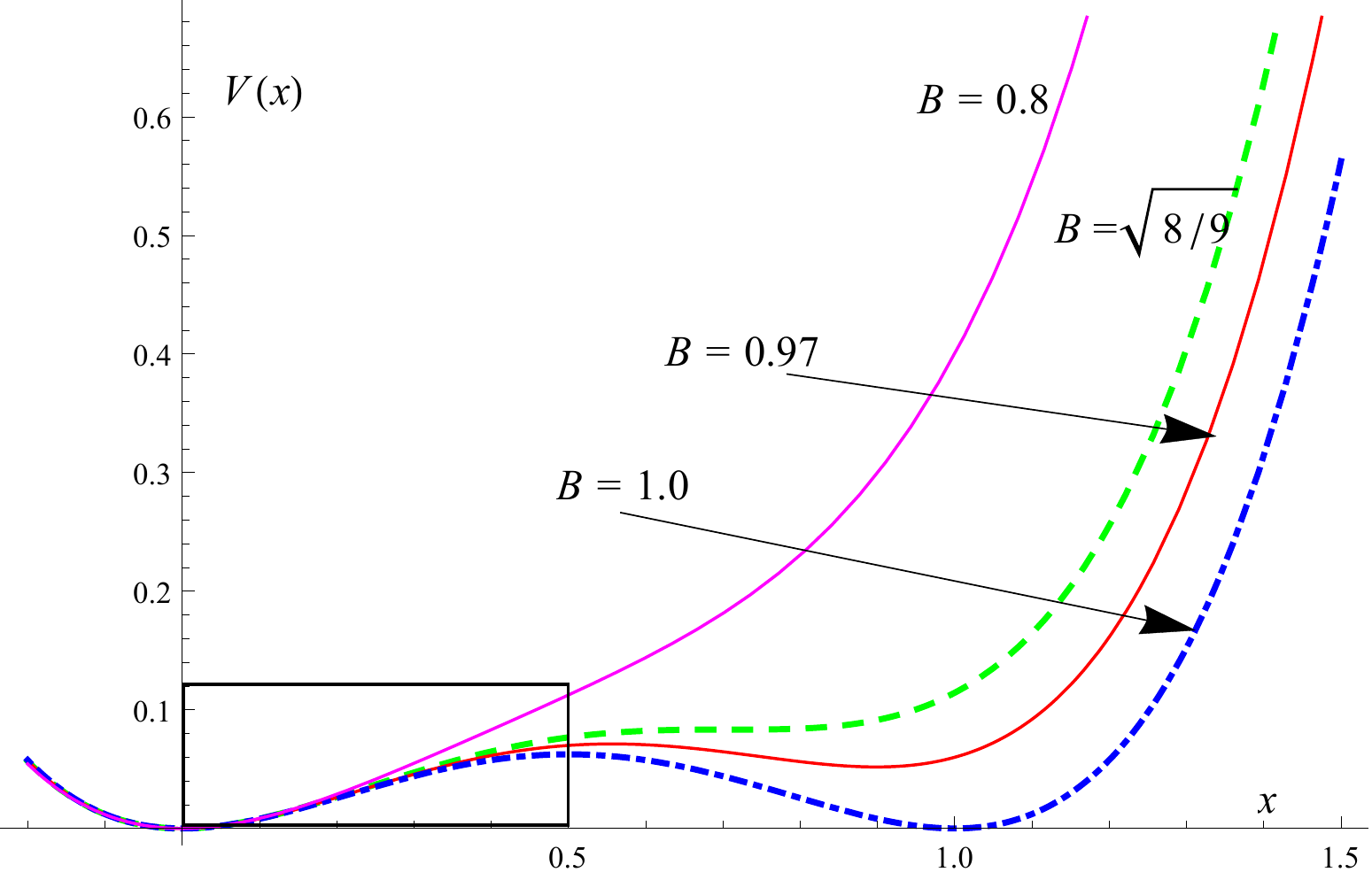}\label{Fig:potentiala}}\\
\subfloat[]{\includegraphics[width=75mm]{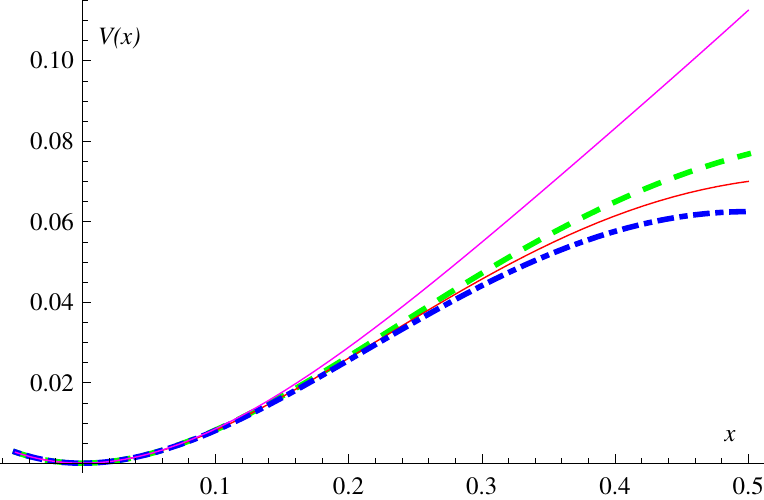}\label{Fig:potentialb}}

\caption{ (color online). \protect\subref{Fig:potentiala} An illustration of the potential for different values of  $B$; \protect\subref{Fig:potentialb} Inset in \protect\subref{Fig:potentiala} zoomed to highlight the behavior as $x \rightarrow 0^{+}$.}\label{Fig:potential}
\end{center}
\end{figure}

\subsection{Inflaton Potential and Background Dynamics}

 On parameterizing the field as $\phi  \rightarrow  x v $, where $x$ is dimensionless, for the chosen potential in \cite{Croon2013}, the inflationary observables reduce to\footnote{In \cite{Croon2013}, a minus sign is missing in the expression for $N$. Further, the power of $(1-x)$ in their Eq.(10), generalised in our Eq.(\ref{Eq:srp3}), should be 4 and not 2.  This error creeps into the numerical calculation  (in table)  also. A summary of standard formulae can be found in \cite{Kolb1990}.}:

\begin{equation}
\epsilon  =\frac{2M_{\text{pl}}^2\left(1+2 x^2-3 x B\right){}^2}{v^2 x^2 \left(1+x^2-2 x B\right)^2}\, ,
\end{equation}

\begin{equation}
\eta  =\frac{2M_{\text{pl}}^2 (1+6 x (x-B))}{v^2 x^2 \left(1+x^2-2 x B\right)}\, ,
\end{equation}

\begin{equation}\label{Eq:srp3}
\xi  = \frac{24 M_{\text{pl}}^4 (2 x-B) \left(1+2 x^2-3 x B\right)}{v^4 x^3 \left(1+x^2-2 x B\right)^2}
\end{equation}
and
\begin{equation}\label{Nex}
 N = -\frac{v^2}{M_{\text{pl}}^2}\int_{x_i}^{x_e} (\frac{V}{V_x}) \, dx \, .
\end{equation}

We point out that the range of possible initial field values is severely restricted if  $B > \sqrt{8/9} $, as the integral for $N$ will not converge on choosing an initial field value beyond the local maximum (see Fig. \ref{fig:integrand}). This rules out the possibility of investigation of deviation from slow-roll or multiphase-inflation. Punctuated inflation \cite{Jain2009} is the simplest and interesting example of multi-phase-inflation which occurs in our model when $ B \sim  \sqrt{8/9} $ and the initial field value is chosen beyond the point of inflection.

We now explore the behavior of the inflationary observables, including $P_s(k)$ and  $f_{\text{NL}}$ as $B$ is
varied from 1 to 0 within the slow-roll regime. Our analysis suggests a broad range of $B$ for which prediction in the $(n_s, r_T)$ plane lies within the 1$\sigma $ level.

\subsection{Discussion about the potential shape}

 The global shape of the potential depends on the coefficient $B$ as shown in Fig. \ref{Fig:potential}. For $ B= 1$, there is one local maximum at $x = 0.5$
and two degenerate minima at $x = 0$ and 1.0. For $B < 1.0$, the minimum at $ x = 1.0$ is lifted. As long as there is such a local maximum, the initial value of the inflaton field should be below the local maximum since otherwise the inflaton would be trapped in the false vacuum.

The false vacuum disappears for $B < \sqrt{8/9}$. Interestingly, if $B$ marginally satisfies the inequality, there appears a flat plateau
at around the point of inflection. If one starts at a suitable value of the field beyond the point of inflection in the above potential, it
is found that one can achieve two epochs of slow roll inflation sandwiching a brief period of departure from inflation (lasting for a little
less than one e-fold ), a scenario which has been dubbed as punctuated inflation \cite{Jain2009}. In fact, it is the point of inflection, around which the potential
exhibits a plateau with an extremely small curvature, which permits such an evolution to be possible.

It needs to be kept in mind that this potential does not provide a single inflation model, but rather a class of the inflationary models as
we vary $B$ from 0 to 1. Fig. \ref{Fig:potential} describes the potential profile for four values of $B$. Interestingly, there are two ways to mimic the quadratic potential case: (a) by changing the
class of potential by decreasing $B$ from 1 to 0  and (b) by taking the initial field value very near to zero (near the first minimum) \cite{Croon2013}. However, 
as shown in Fig. \ref{Fig:potentialb}, each member of the class is asymptotic to the quadratic case if $ x_i <  0.12$ irrespective of the value of $B$.

\begin{figure}[H]
\begin{center}
\subfloat[]{\includegraphics[height=60mm,width=70mm]{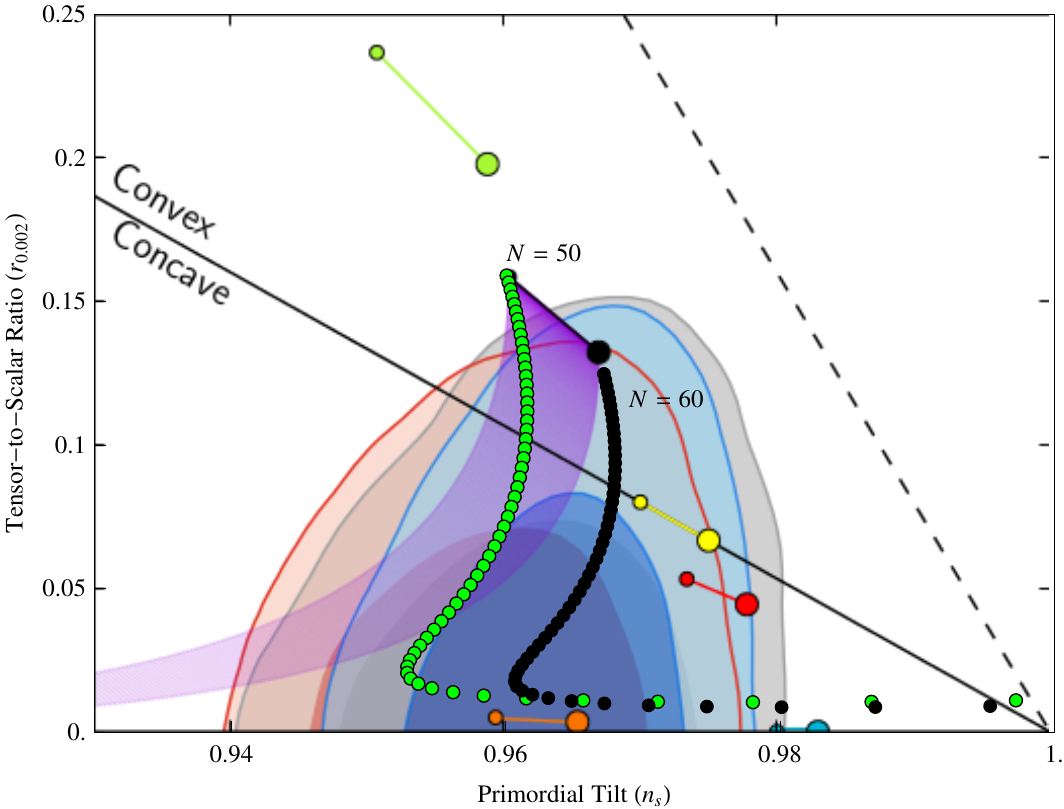} \label{F:nakayamaa}}\\
\subfloat[]{\includegraphics[height=60mm,width=70mm]{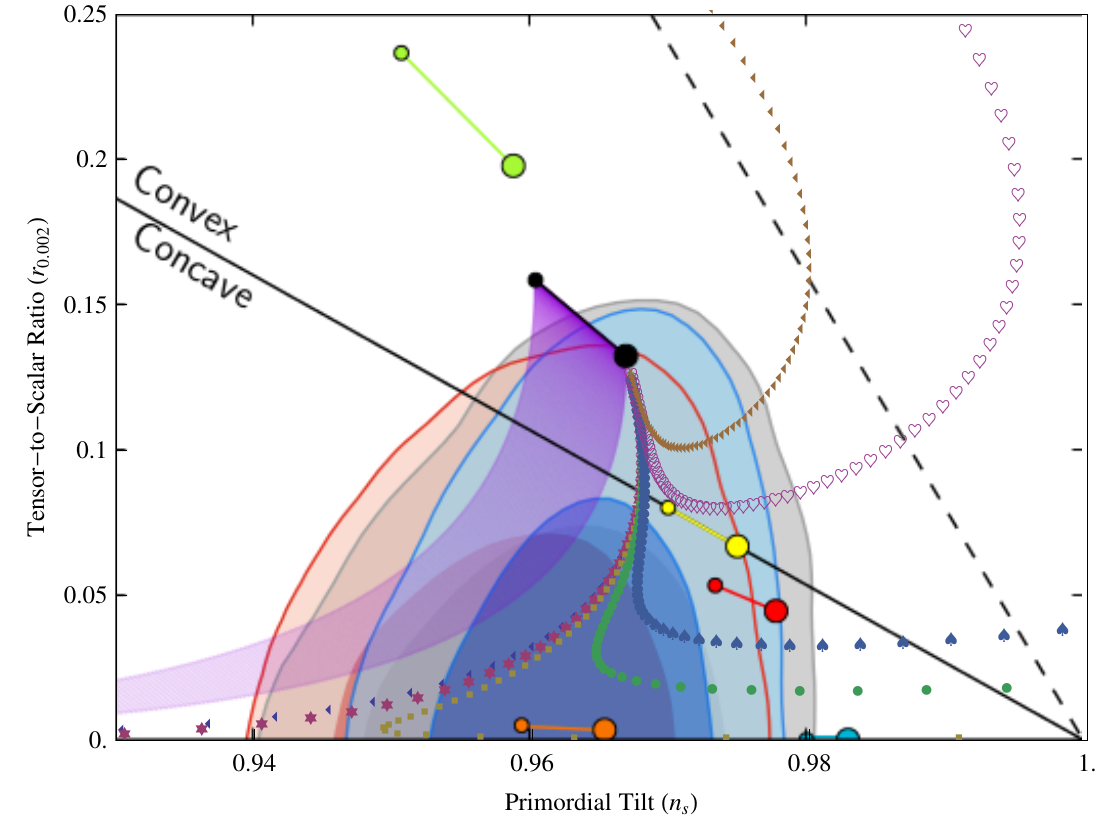} \label{F:nakayamab}}

\caption{ (color online). The reproduction of the work by Nakayama et al. \cite{Nakayama2013}. \protect\subref{F:nakayamaa} The  two curves correspond to $N$ = 50 (green) and 60 (black). Each curve is generated by varying $x_i$ from 0.005 to 1.0 in steps of 0.012. In this plot we have fixed   $B =0.93$ for both curves. \protect\subref{F:nakayamab} The seven curves correspond to B: 0.96, 0.95, 0.94, 0.92, 0.90, 0.80, 0.70 for   $N$ = 60.   Each curve is again generated by varying $x_i$ as in \protect\subref{F:nakayamaa}.  \label{F:nakayama}}
\end{center}
\end{figure}

\begin{figure}[H]
\centering

\subfloat[]{\includegraphics[height=50mm,width=70mm]{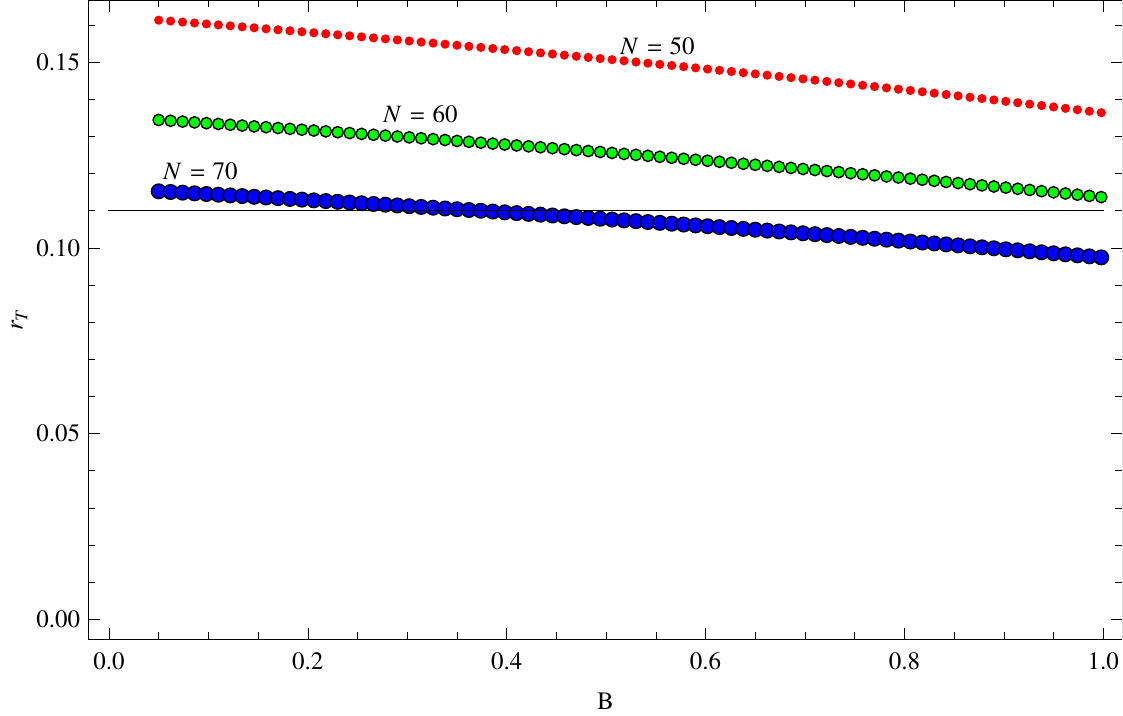}\label{fig:xi1a}}\\
\subfloat[]{\includegraphics[height=50mm,width=70mm]{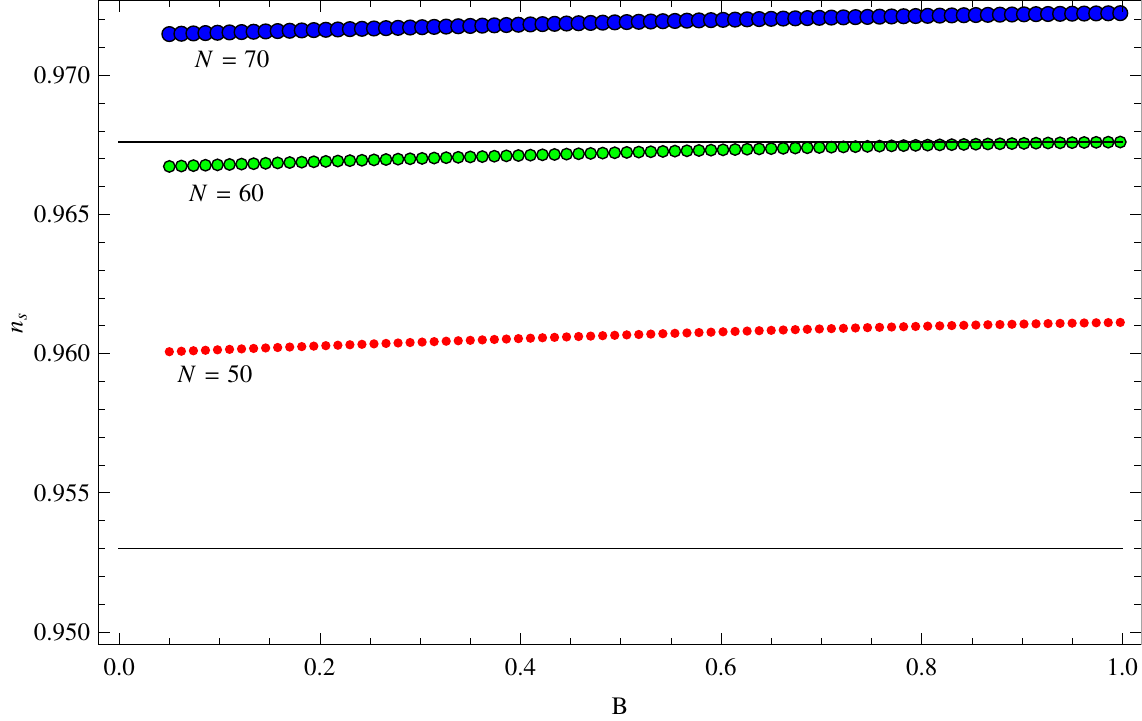}\label{fig:xi1b}}\\

\subfloat[]{\includegraphics[height=55mm,width=75mm]{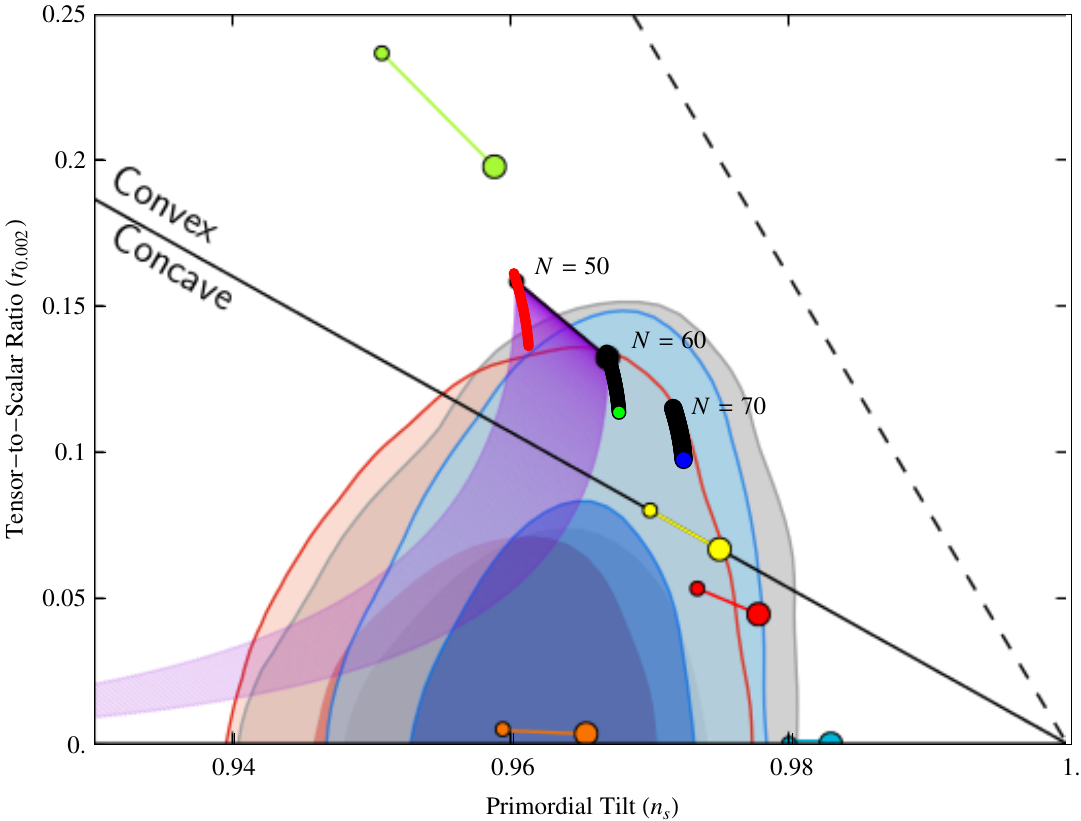}\label{fig:xi1c}}
\caption{ (color online). \protect\subref{fig:xi1a} The tensor-to-scalar perturbation ratio $r_T$ as a function of \textit{B} for $N = 50$ (red), 60 (green) and 70 (blue). In this plot we have taken   $x_i =0.1$ for all three curves. The  horizontal line shows the upper limit for $r_T = 0.11$. \protect\subref{fig:xi1b} spectral index $n_s$ as a function of \textit{B} with same color coding as in Fig. \protect\subref{fig:xi1a}. Two horizontal lines show the constraint on $n_s$, i.e., $ n_s = 0.9603 \pm  0.0073$. \protect\subref{fig:xi1c} Prediction in the $(n_s, r_T)$ plane corresponding to \protect\subref{fig:xi1a} and \protect\subref{fig:xi1b}. For example, each red point in the $(n_s, r_T)$ plane corresponds to a value of  $r_T$ ( red point in \protect\subref{fig:xi1a})  and a value of  $n_s$ (red point in \protect\subref{fig:xi1b}), both calculated for a particular value of $B$.\label{fig:xi1} }

\end{figure}

\begin{figure}[H]
  \begin{center}
 
  \subfloat[]{\includegraphics[height=50mm,width=70mm]{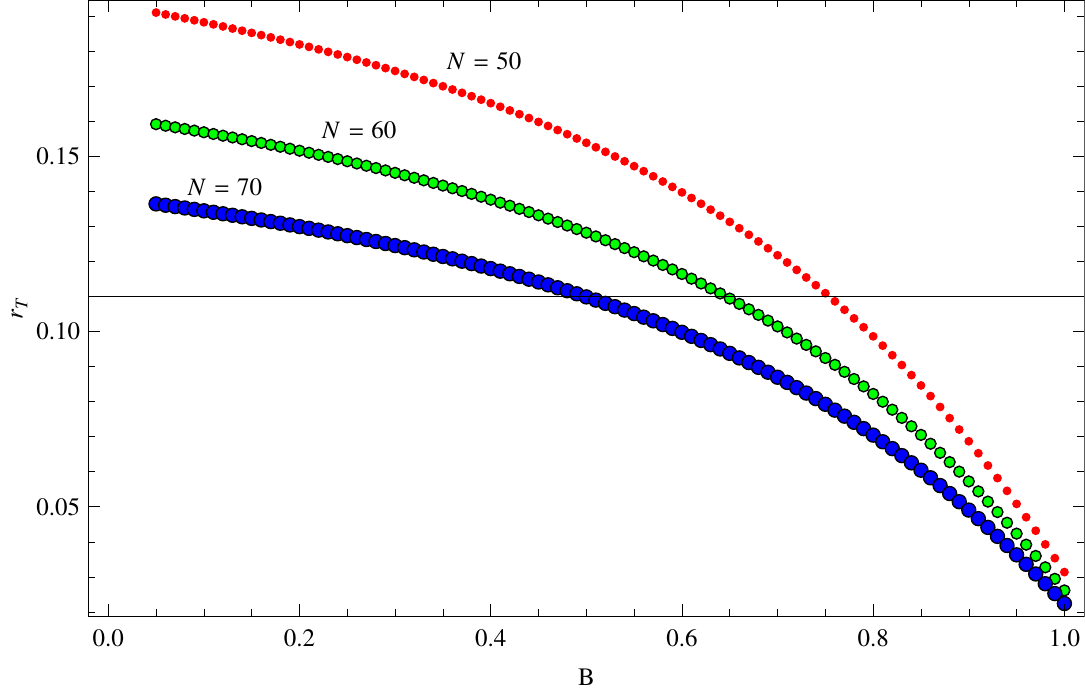}\label{fig:xi4a}} \\
  \subfloat[]{\includegraphics[height=55mm,width=75mm]{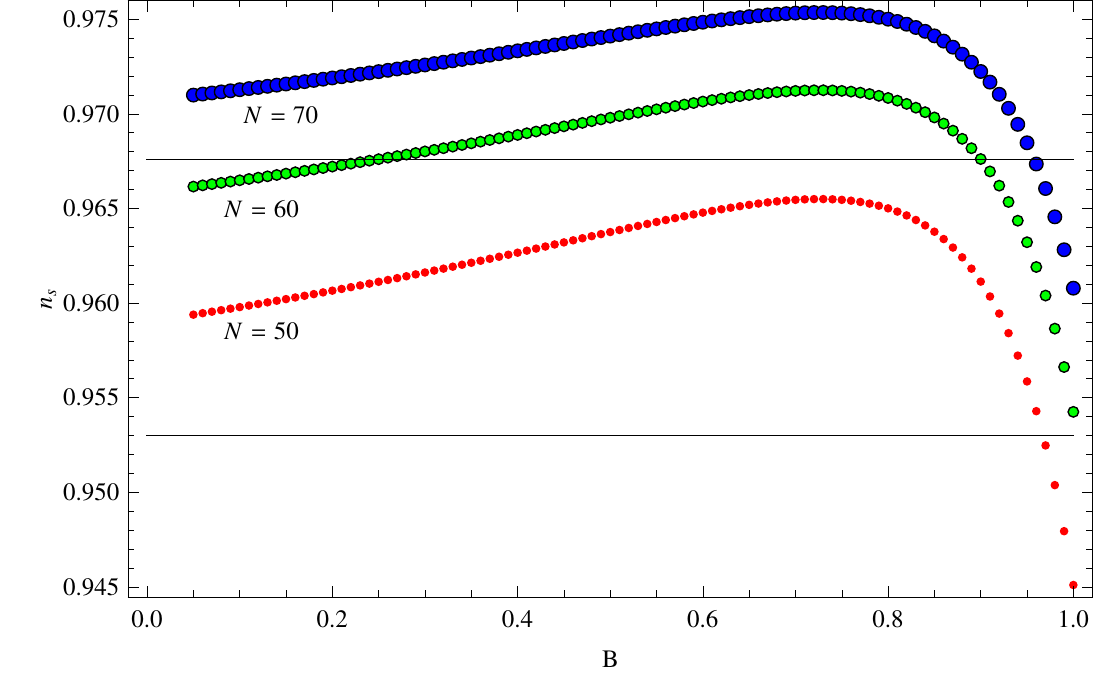}\label{fig:xi4b}}\\
  
  \subfloat[]{\includegraphics[height=55mm,width=75mm]{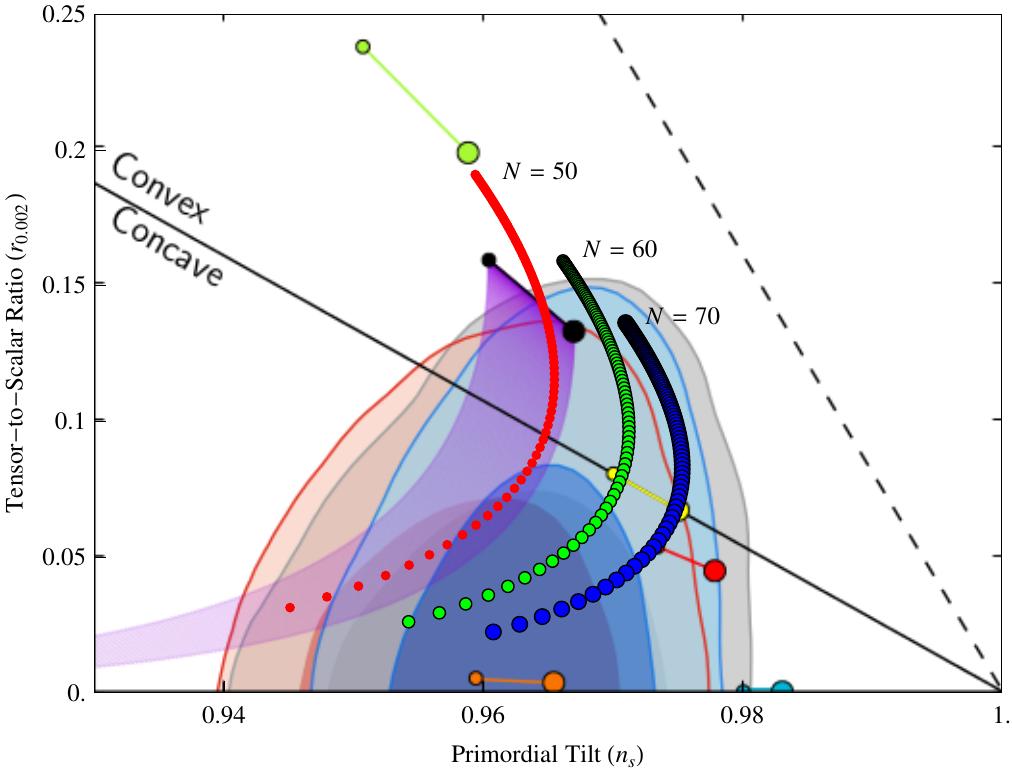}\label{fig:xi4c}}
 
  \caption{ (color online). Same as in Fig. \ref{fig:xi1}, but here we have taken   $x_i =0.4$ (near the local maximum) for all three curves. Note the  characteristic interesting behavior when field starts near the local maximum.\label{fig:xi4}}
  \end{center}
  \end{figure}

\section{ Numerical Results}

 We set out to explore the model for the entire range of $B$ values from 0 to 1 and for an arbitrary initial field value $\phi_i$ (or equivalently $x_i$). We need to keep in mind that an arbitrary large initial field value can be taken only for those members of the class for which $B < \sqrt{8/9}$, otherwise either deviation from slow roll will occur (when $ B = \sqrt{8/9}$ ) or the field will be trapped in the false vacuum (when $B > \sqrt{8/9}$ ).
 
 \begin{figure}[H]
 \begin{center}
 \subfloat[]{\includegraphics[height=60mm,width=75mm]{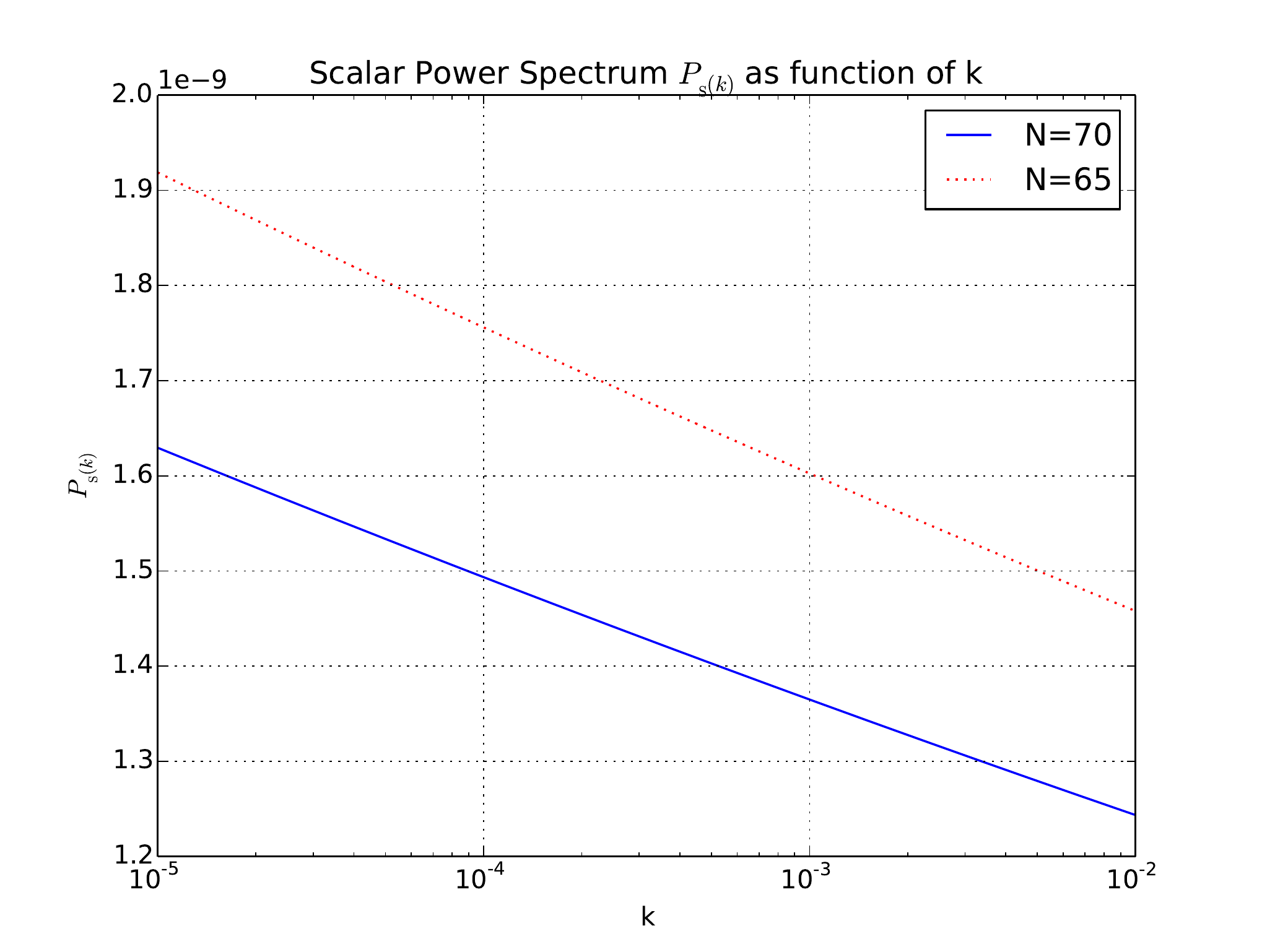}\label{fig:powerspectrum}}\\
 \subfloat[]{\includegraphics[height=60mm, width=75mm]{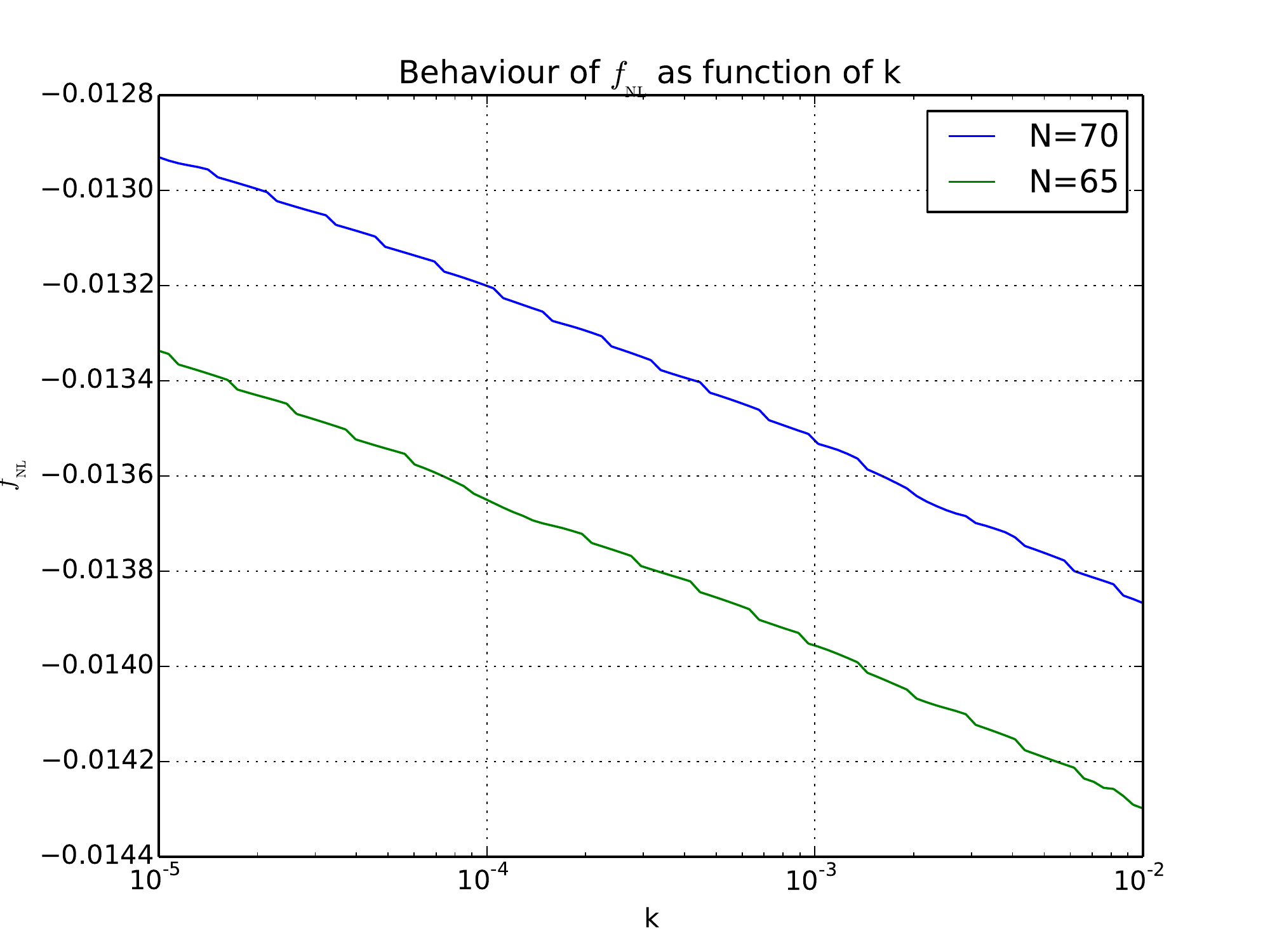}\label{fig:fnl}}
 \quad
 
 \caption{ (color online). \protect\subref{fig:powerspectrum} The scalar power spectrum as a function of $k$  plotted on a logarithmic scale for $B = 0.96$ and $x_i = 0.4$ over two e-folds. \protect\subref{fig:fnl} A plot of $f_{\text{NL}}$ as a function of $k$ on a logarithmic scale. The values of $B$ and $x_i$ are the same as in \protect\subref{fig:powerspectrum}.}

 \end{center}
 \end{figure}

\subsection{Prediction in $(n_s, r_T)$ plane}

Fig. \ref{fig:xi1} exhibits how our analysis differs from the past work of Nakayama et al. \cite{Nakayama2013} (see Fig. \ref{F:nakayama}), where effects mainly due to  varying initial field value are considered. In contrast, we have varied $B$ continuously for each fixed initial field value---from the one near the minimum at $x = 0$ to the other near the local maximum---and for three e-fold values: N = 50, 60 and 70. However, we have shown the analysis only for the two extreme initial field values because of clarity and reasons of space (see Figs. \ref{fig:xi1} and \ref{fig:xi4}).

  Plots only on the $(n_s, r_T)$ plane, as done in \cite{Nakayama2013}, hide valuable information of initial conditions for which the model lies within 1$\sigma$. Notice that very near to $x= 0.005$, predictions (i.e., points in the $(n_s, r_T)$ plane)  start from the quadratic potential (the black line), found concordant with the Planck data,  and then graze the contour as $x$ is increased; see Figs. \ref{F:nakayama} and \ref{fig:Plancknsr}.

From Fig. \ref{fig:xi1a}, it is clear that for $N$ = 50 (red) and 60 (green), this model is completely above the bound imposed by Planck ($r_T < 0.11$) for the entire range of $B$---here the field starts near the origin; whereas for $N$ = 70 (blue), the model is at the margin of this bound.

 When we contrast the above situation with  Fig. \ref{fig:xi4a}, we find that the upper limit on $r_T$ is satisfied for sufficiently large values of $B$ (red and green curves fall below the horizontal line for  $B \gtrsim 0.75$ and $B \gtrsim 0.65$ respectively). It  is emphasized that this is the case when the field starts near the local maximum ($x \sim 0.4$). Fig. \ref{fig:xi4c} makes this explicit, showing that even a part of the red curve ($N$ = 50) is within 1$\sigma$ of the Planck value. Although there is no appreciable disagreement between Planck and BICEP2 results as clarified in \cite{audrennote2014}, a slightly larger tensor-to-scalar ratio $(r_T \sim 1.6)$, as required by BICEP2, is within reach (e.g., when $B \sim 0.3$) as shown in Fig. \ref{fig:xi4a}. Another way to realize a large $r_T$ is by assuming the non-monotonicity of the slow-roll parameter $\epsilon$, as considered in a model somewhat similar to ours \cite{Choudhuryaccurate2014}. 

\subsection{Non-Gaussianity}

Non-Gaussianity is  now a standard cosmological observable, comparable to the spectral index ($n_s$) and tensor-to-scalar ratio $r_T$, and is a powerful
discriminant between competing models. It is therefore desirable to study aspects of non-Gaussianity for this model too. We have numerically calculated the equilateral limit of the local $f_{\text{NL}}$ parameter \cite{Hazra2013}. In Fig. \ref{fig:powerspectrum},
the power spectrum is plotted as a function of $k$, from which it is clear that the
spectrum is not scale invariant on any scale (otherwise it would have been a horizontal straight line). In Fig. \ref{fig:fnl}, the non-Gaussianity parameter  $f_{\text{NL}}$ are plotted as a function of $k$, from which we observe that
the very low values of $|f_{\text{NL} }| \,\, (\sim 0.01)$ lie in the range predicted by Planck data for local $f_{\text{NL}}$ = 2.7 $\pm $ 5.8 \cite{Collaboration2013}. The very small non-Gaussianity, inferred from \mbox{BICEP2} data owing to comparatively larger $r_T$ \cite{damiconongaussianity2014} (see also \cite{Choudhuryreconstructing2014}), is compatible with our results also  \cite{hazrawhipped2014}. This observation is also consistent with the fact that single-field slow roll models predict a very small non-Gaussianity  \cite{Maldacena2003}.

\section{Summary and Conclusion}

In this Letter, we have explored in detail the $(B, \phi_{i})$ space within the slow-roll regime for the quartic potential. We have explicitly shown
that on varying $B$, interesting patterns on the variation of $n_s$ and $r_T$ arise \textit{only} when field dynamics captures
the deviation from the quadratic case, i.e., the field should start near the local maximum, not near the origin. Even though this parameter space has been
moderately studied in \cite{Nakayama2013}, the authors  restricted their analysis to a few discrete values of $B$. Similarly Ellis et al. \cite{ellisexploring2014} have carried out a statistical study with just four explicit values of numerical coefficient in a two field model to explore concordance with the preferred observational values of $n_s$ and  $r_T$.

One of the main well-known problems---constraints from Planck on the upper limit on $r_T$---has been reanalyzed here and we find that there is no respite for the model \textit{solely} due to removing degeneracy between the vacua. On the other hand, merging variations of $B$ with the field starting \textit{near} the local maximum, the model's predictions for $(n_s, r_T)$ lie well within 1$\sigma $ for a large part of the $(B, \phi_i )$ space (see Fig. \ref{fig:xi4}). Further, our Fig. 4a clearly shows that  large $r_T$ can be generated when the potential is non-degenerate and the field starts near the local maximum. On non-Gaussianity aspects ( which were not considered in previous works \cite{Croon2013},\cite{Nakayama2013}), we have explicitly calculated the local $f_{\text{NL}}$ parameter for each mode of our
interest and verified that this model indeed satisfies the constraint imposed by Planck data on non-Gaussianity \cite{Collaboration2013}, and is compatible with BICEP2 measurements.

\section*{Acknowledgments}

RK thanks Dheeraj Kumar Hazra for useful communication and Tarun Choudhari for assistance with Latex. We are grateful to Rudnei O. Ramos and Sayantan Choudhury for pointing out their works \cite{ramospower2013} and \cite{Choudhuryaccurate2014, Choudhuryreconstructing2014} respectively. We also thank an anonymous reviewer whose comments helped us highlight the subtle points of our study. The work of RK was partially supported by CSIR, New Delhi with the award of JRF (No. 09/045(0930)/2010-EMR-I).

\section*{Appendix}

\begin{figure}[H]
\begin{center}
\subfloat[]{\includegraphics[width=55mm]{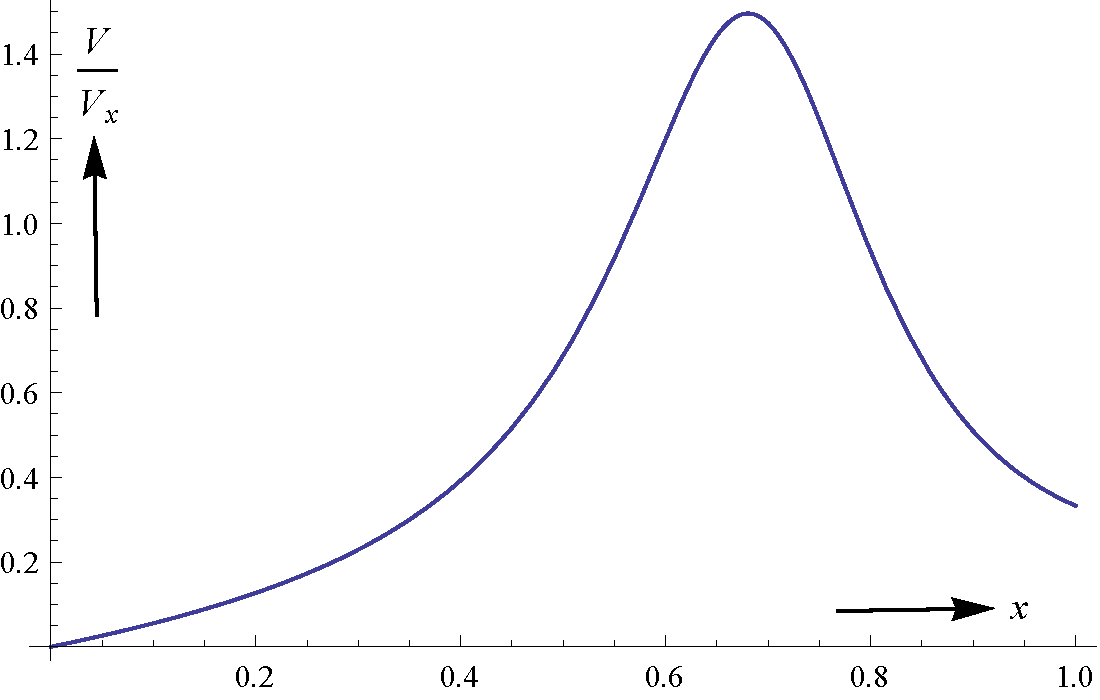}\label{fig:integranda}}\\
\subfloat[]{\includegraphics[width=55mm]{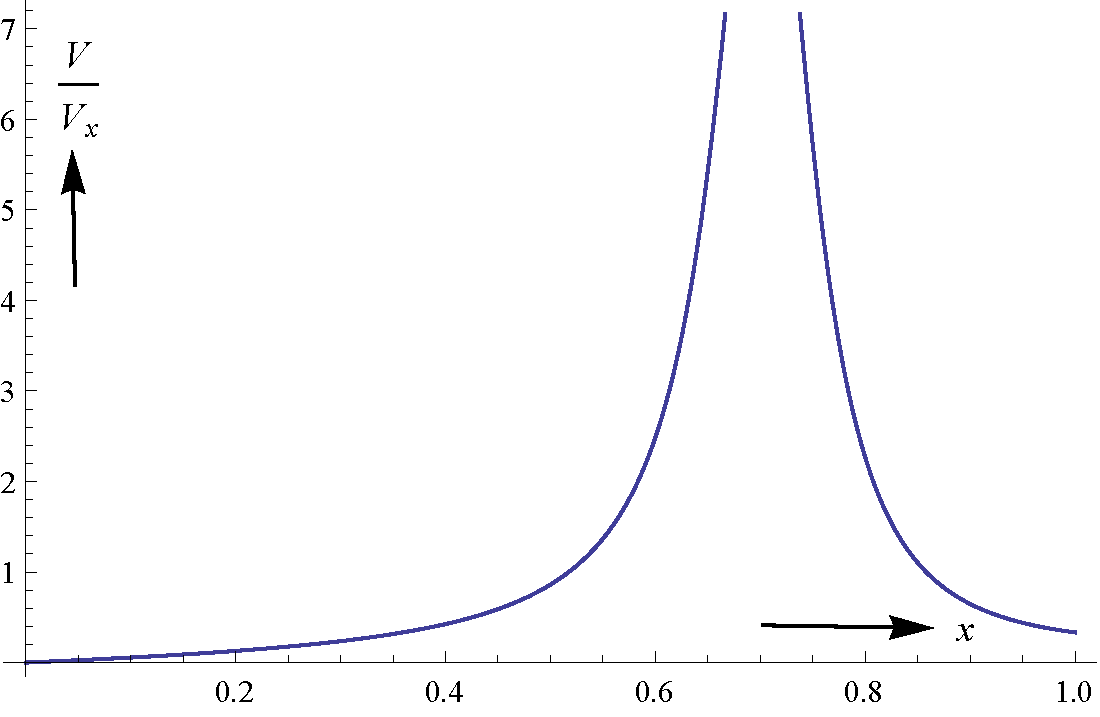}\label{fig:integrandb}}\\
\subfloat[]{\includegraphics[width=55mm]{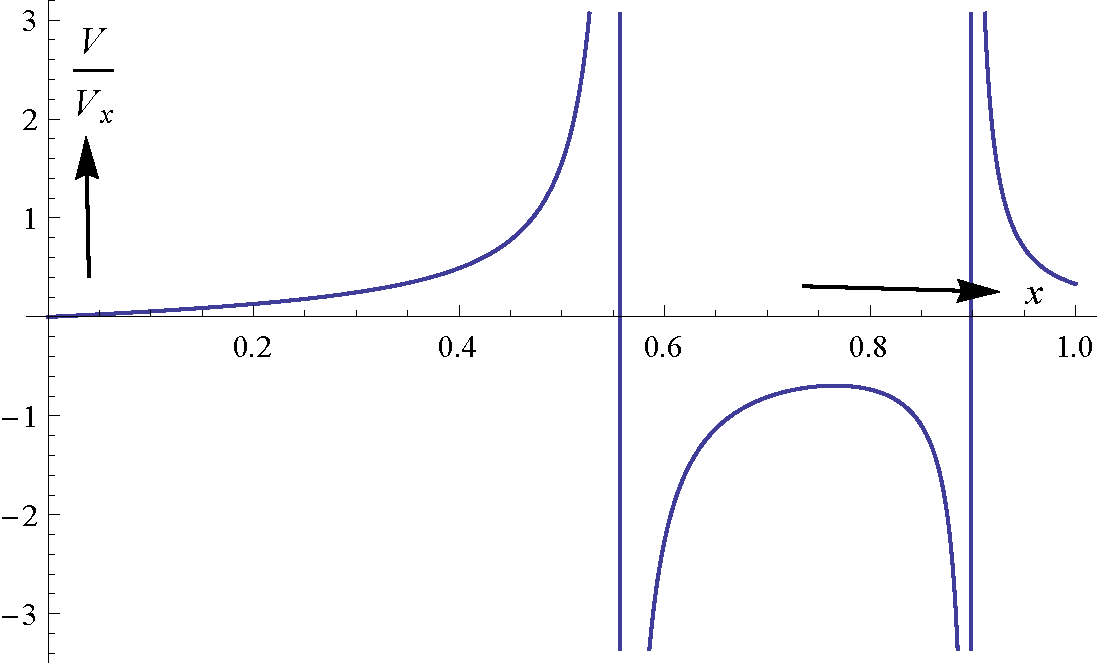}\label{fig:integrandc}}
\caption{ Schematic plot of integrand for $N$ for \protect\subref{fig:integranda} $B = 0.92$, \protect\subref{fig:integrandb} $B = 0.94$  and \protect\subref{fig:integrandc} $B = 0.97$; see Eq. \ref{Nex}. \label{fig:integrand}}
\end{center}
\end{figure}

In Fig. \ref{fig:integrand}, we plot the integrand for $N$ for 3 values of $B$. (As already mentioned, one minimum disappears if $B < \sqrt{8/9} = 0.9428$.)  We can clearly see how a discontinuity in $\frac{V}{V_x}$ develops as $B$ increases above $\sqrt{8/9}$. One cannot take any point on the $x$ axis beyond the local maximum, which is near $x =0.5$ with a slight variation as we change $B$ (see Fig. \ref{Fig:potential}).

\begin{figure}[H]
 \begin{center}
 \subfloat{\includegraphics[height=50mm,width=80mm]{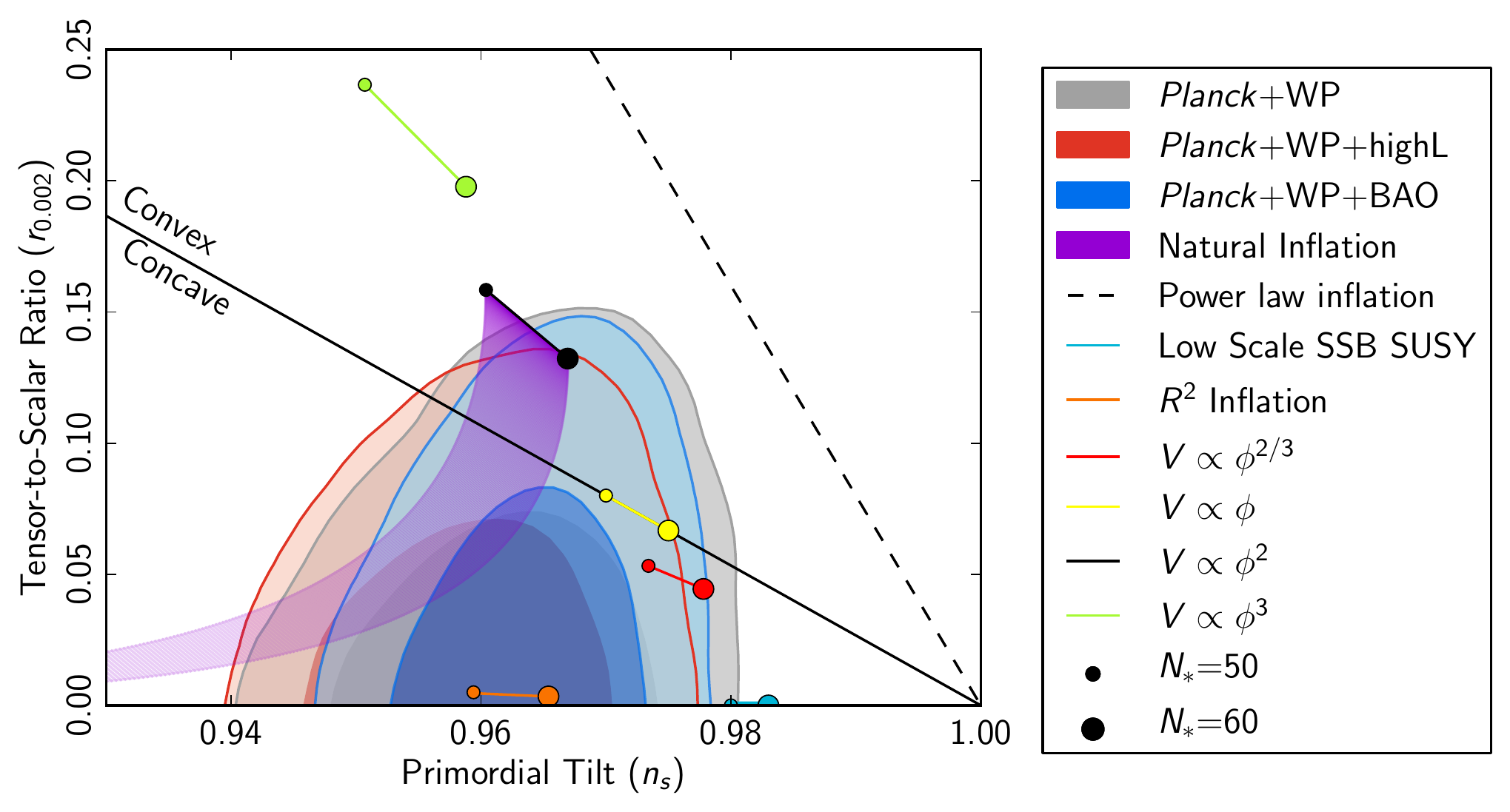}}
 \caption{ (color online). Marginalized joint 68\% and 95\%~CL regions for $n_\mathrm{s}$ and $r_{0.002}$. Also shown are observational $1\sigma$ (dark) and $2\sigma$ (light) constraints from the Planck satellite \cite{Ade2013}:
 Planck + WMAP polarization (gray), Planck + WMAP polarization + high-$\ell$ CMB measurement (red),
 Planck + WMAP polarization + baryon acoustic oscillation (blue).
 Filled circles connected by line segments show the predictions from chaotic inflation with $V \propto \phi^3$ (green), $\phi^2$ (black), $\phi$ (yellow), $\phi^{2/3}$ (red) and $R^2$ inflation (orange),
 for $N=50$ (small circle)--$60$ (big circle). The purple band shows the prediction of natural inflation \cite{Ade2013}\label{fig:Plancknsr}.}
 \end{center}
 \end{figure}

 Fig. \ref{fig:Plancknsr} is from Planck Data on which we superimpose our predictions---following \cite{Croon2013} and \cite{Nakayama2013}.
 
\section*{References}

\bibliographystyle{elsarticle-num}
\bibliography{elspoly}

 \end{document}